\documentclass[12pt,a4paper,final]{iopart}

\usepackage{iopams} 
\usepackage{lipsum}
\usepackage[numbers]{natbib}
\def\newblock{\hskip .11em plus .33em minus .07em} 
\usepackage{graphicx,caption}

\usepackage{relsize} 
\newcommand{\AFLOWpi}{ {\sf AFLOW$\mathlarger{\mathlarger{\mathlarger{{\pi}}}}$}}

\begin{document}

\title[]{Improved electronic structure and magnetic exchange interactions in transition metal oxides}

\author{Priya Gopal$^{1}$, Riccardo De Gennaro$^{2}$, Marta S. Gusmao$^{3}$, Rabih Al Rahal Al Orabi$^{1}$,  Haihang Wang$^{5}$, Stefano Curtarolo$^{4}$, Marco Fornari$^{1,4}$ and Marco Buongiorno Nardelli$^{4,5}$}

\address{$^{1}$ Department of Physics, Central Michigan University, Mt. Pleasant, MI 48859 }
\address{$^{2}$ Dipartimento di Fisica, Universit\`a� di Roma Tor Vergata, Via della Ricerca Scientifica 1, 00133 Roma, Italy}
\address{$^{3}$Department of Physics, Federal University of Amazonas, Manaus, Amazonas, Brazil}
\address{$^{4}$ Center for Materials Genomics, Duke University, Durham, NC 27708, USA}
\address{$^{5}$ Department of Physics, University of North Texas, Denton, TX 76203, USA}
\ead{mbn@unt.edu}
\vspace{10pt}
\begin{indented}
\item[]
\end{indented}

\begin{abstract}
We discuss the application of the \underline{A}gapito \underline{C}urtarolo  and \underline{B}uongiorno \underline{N}ardelli (ACBN0) pseudo-hybrid Hubbard density functional to several transition metal oxides. ACBN0 is a fast, accurate and parameter-free alternative to traditional DFT+\textit{U} and hybrid exact exchange methods. In ACBN0, the Hubbard energy of DFT+\textit{U} is calculated via the direct evaluation of the local Coulomb and exchange integrals in which the screening of the bare Coulomb potential is accounted for by a renormalization of the density matrix. 
We demonstrate the success of the ACBN0 approach for the electronic properties of a series technologically relevant mono-oxides  (MnO, CoO, NiO, FeO, both at equilibrium and under pressure). We also present results on two mixed valence compounds, Co$_3$O$_4$ and Mn$_3$O$_4$. Our results, obtained at the computational cost of a standard  LDA/PBE calculation, are in excellent agreement with hybrid functionals, the GW approximation and experimental measurements.

%
%
%
%
%

\end{abstract}


\section{Introduction}
\label{sec:intro}
Density functional theory (DFT) combined with local (LDA) or semi-local approximations (GGA) \cite{perdew1981self,perdew1996generalized} has become a standard tool for performing electronic structure calculations of materials. Despite its capabilities in predicting many physical properties, the method fails both quantitatively and qualitatively in strongly correlated (SC) electron systems. This inaccurate description is the consequence of the incorrect treatment of the exchange interaction since the approximations do not sufficiently cancel the electron self-interaction \cite{perdew1981self,jones1989density}. Even in simple transition metal (TM) monoxides the failures of the local/semi-local approximation is dramatic. In MnO and NiO, for instance, LDA/GGA predicts a very small band-gap (up to 80 \% smaller than experiments) while both CoO and FeO are incorrectly described as metallic and ferromagnetic \cite{anisimov1997first}. Experimentally, all the four TM monoxides are well known charge transfer insulators and robust antiferromagnets (AFM)  with significant magnetic exchange interactions and N\'eel ordering temperatures (transition from a paramagnetic to an antiferromagnetic ordering) \cite{roth1958magnetic}. Similar failures of the LDA/GGA approach are also seen in Mn$_3$O$_4$ and Co$_3$O$_4$ in which the same TM atom is found in two different oxidation states \cite{chartier1999ab,roth1964magnetic}.

Numerous schemes to correct the self-interaction error have been proposed: self-interaction corrected (SIC) LDA~\cite{perdew1981self}, variational pseudo-SIC \cite{filippetti2011variational}, DFT+\textit{U}  \cite{anisimov1997first,himmetoglu2014hubbard,toroker2011first}, inclusion of fraction of exact exchange with different screenings like the exact exchange methods (EXX-OEP) \cite{engel2009insulating}, screened exchange LDA (sX-LDA) \cite{gillen2013accurate}, hybrid functionals such as HSE \cite{heyd2003hybrid} and PBE0 \cite{perdew1996rationale}, multiplicative potentials for localized basis sets (modified Becke-Johnson potential (mBJ) \cite{tran2009accurate}), and the quasiparticle GW methods \cite{aryasetiawan1998gw}. Among the different approaches, the simpler and more used method is the DFT+\textit{U}, where the spurious intra-atomic electron-electron interaction within selected sub-shells is treated by semiempirical Coulomb and exchange integrals. One of the major disadvantage of the DFT+\textit{U} method is the ambiguity in the choice of \textit{U}. Conventionally, the \textit{U} is treated as an empirical parameter and fitted to either experiments or higher order functionals. Since this approach is semi-empirical in nature, it limits the predictive power of the calculations. The \textit{ab initio} determination of \textit{U} through constrained LDA (cLDA) or linear response methods \cite{gunnarsson1989density2,cococcioni2005linear}  requires supercell calculations and can become impractical in large-scale simulations. 

Some of us have recently introduced a new pseudo-hybrid Hubbard density functional, the \underline{A}gapito \underline{C}urtarolo  and \underline{B}uongiorno \underline{N}ardelli  functional (ACBN0),  as a fast, accurate, and parameter-free alternative to traditional DFT+\textit{U} and hybrid exact exchange methods \cite{agapito2015reformulation}.
In ACBN0, the Hubbard energy of DFT+\textit{U} is calculated via the direct evaluation of the local Coulomb and exchange integrals in which the screening of the bare Coulomb potential is accounted for by a renormalization of the density matrix. Through this procedure, the values of \textit{U}, defined as the difference of the Coulomb and exchange integrals, are thus functionals of the electron density and depend directly on the chemical environment and crystalline field. ACBN0 satisfies the rather ambitious criteria outlined by \cite{pickett1998reformulation} in one of the first seminal articles on LDA+\textit{U}: i) ACBN0, reduces to LDA/GGA when LDA/GGA is known to be good; ii) the energy is given as a functional of the density; iii) the method specifies how to obtain the local orbital in question; iv) the definition of the Coulomb and exchange integrals is provided unambiguously; and v) the method predicts antiferromagnetic insulators when appropriate and improves the description of highly correlated metals. ACBN0 corrects both the band gap and the relative position of the different bands, in particular the ones deriving from the \textit{d} orbitals of TM atoms.
We have tested ACBN0 for several systems (semiconductors and nitrides \cite{supka2017aflow}, Zn and Cd oxides and chalcogenides \cite{gopal2015improved}) and showed that it is at par with other advanced functionals like the HSE06 and SIC functionals in accuracy, while it outperforms them in computational efficiency. In this work, we demonstrate the success of ACBN0 in predicting the electronic and magnetic exchange interactions in MnO, FeO, CoO, NiO, Co$_3$O$_4$ and Mn$_3$O$_4$. 

This article is organized as follows: in Sec.\  \ref{sec:methods}, we discuss the computational methods; the electronic and the magnetic properties of  the  mono-oxides are presented in Sec. \ref{sec:resultsTMmono}; and the mixed-valence oxides Co$_3$O$_4$ and Mn$_3$O$_4$ are discussed in Sec.\  \ref{sec:resultsMV}. 

\section{Methodology}
\label{sec:methods}

ACBN0 is based on the DFT+\textit{U} energy functional as formulated by \cite{dudarev1998electron}:
\begin{equation*}
E_{\rm{DFT}+\it{U}}=E_{\rm{DFT}} + E_{\it U} -E_{DC}
\end{equation*}
where $E_{\rm{DFT}}$ is the DFT energy calculated using a LDA or GGA functional; $\textit{U}$ is an effective on-site Coulomb interaction given by the difference between the Coulomb and exchange integrals  and $E_{DC}$ takes care of the double counting terms in the energy expansion. For a complete discussion on the foundations of ACBN0, we remand the reader to Ref. \cite{agapito2015reformulation,gopal2015improved}. We compute on-the-fly the local Coulomb  and exchange  integrals for a specific orbital manifold \textit{via} a self-consistent procedure based on an \textit{ad hoc} renormalization of the density matrix. The value of \textit{U} is specific for the material and for the chemical environment and the crystalline field. For instance, in mixed-valence systems the method can naturally distinguish between the different oxidation states of each chemically equivalent element  in the material (see Sec. \ref{sec:resultsMV}). 

The ACBN0 procedure is implemented using a projection on atomic orbitals \cite{agapito2013effective, agapito2016accurate, agapito2016accurate2} in the high-throughput framework \AFLOWpi\ that provides automatic workflows for the calculation of the dielectric constant, the phonon spectra, and diffusive transport coefficients \cite{supka2017aflow}.
Our implementation of ACBN0 uses norm-conserving pseudo-potentials. In order to verify the transferability of the $U$ values, we have tested the electronic structure properties using different pseudo-potentials (ultrasoft \cite{garrity2014pseudopotentials} and PAW) and also all electron methods. 
All the DFT calculations in this paper are carried out by using the Perdew-Zunger (PZ) or Perdew-Burke-Ernzerhoff (PBE) functionals as starting point. We used the QE package \cite{giannozzi2009quantum} as electronic structure engine in  \AFLOWpi. For the ACBN0 calculations, we used scalar-relativistic norm conserving  pseudo-potentials from the PSlibrary 1.0.0 \cite{dal2014pseudopotentials}. The energy cut-off for the electronic convergence is set to 350 Ry (60 Ry) for the norm-conserving pseudopotentials (ultrasoft) pseudo-potentials. 

As a further validation of our results, we performed all electrons calculation using the full-potential linearized augmented plane wave approach as implemented in the WIEN2K code \cite{blaha2001wien2k} with  the modified Becke-Johnson (mBJ) functional \cite{tran2009accurate}. The muffin-tin radii (R$_{MT}$) were chosen small enough to avoid overlapping during the optimization process. A plane wave cutoff corresponding to R$_{MT}$K$_{max}$ = 7 was used in all calculations. The radial wave functions inside the non-overlapping muffin-tin spheres were expanded up to $l_{max}$ =12. The charge density was Fourier expanded up to $G_{max}$ =16 \AA$^{-1}$. Total energy convergence was achieved with respect to the Brillouin zone (BZ) integration mesh with 500 $k$-points. 

The exchange coupling constants $J_{ij}$ were extracted from the total-energy differences of a number of different magnetic structures mapped onto a nearest- and the next-nearest-neighbor Heisenberg Hamiltonian \cite{fischer2009exchange}. Based on the calculated $J_{ij}$, the transition temperature (T$_N$) was obtained using classical Monte Carlo (MC) simulations. 
MC simulations have been run for a face centered cubic lattice representing the TM atoms in the rocksalt structure. To minimize finite size effects, we chose a $20\times20\times20$ lattice with a simulation time of 10000 steps  for the Metropolis algorithm \cite{bhanot1988metropolis}.

\section{Results: TM  mono-oxides}
\label{sec:resultsTMmono}

The \textit{3d} TM monoxides (MnO, FeO, CoO and NiO) crystallize in the rock salt structure (B1, Fm3m, space group 225) \cite{roth1958magnetic,mehl2016aflow} and exhibit antiferromagnetic ordering of type II (AF2) below their N\'eel temperature ($T_N$). The AF2 order along the [111] direction reduces the symmetry to a rhombohedral one (space group \textit{R3m}, number 166) containing two formula units (4 atoms).  
We use this structure to calculate all the structural and electronic properties. The small deviations from the ideal cubic lattice were ignored in our calculations.  For a schematic representation of the structures see Ref. \cite{anisimov1997first}. 

The \textit{U} values for the starting PBE and LDA equilibrium volumes are tabulated in Table \ref{tab:Ueff-tmo}. The $U$ on the TM atom increases  from Mn to Ni consistently with the \textit{d} orbitals occupancy and with the increased degree of localization. 
The oxygen \textit{2p} values also vary slightly across the \textit{3d} row indicating the dependence of \textit{U} from the chemical environment.

\begin{table}
\footnotesize
\centering
\captionsetup{width=\linewidth}
\caption{\label{tab:Ueff-tmo} Converged values of the effective on-site Coulomb parameter $U$ (in eV) for the transition metal (TM) $3d$ and the oxygen $2p$ states.}
\begin{tabular}{lcccc}
\mr
& MnO  &FeO & CoO & NiO \\
\hline
\bf{PBE} & & & & \\
TM-$3d$      & 4.67 & 5.73  & 6.38 & 7.63 \\
Oxygen $2p$  & 2.68 & 2.88  & 2.60 & 3.00 \\
\hline
\bf{LDA} & & & & \\
TM-$3d$      & 4.72 & 5.82  & 6.45 & 7.49 \\
Oxygen $2p$  & 2.55 & 2.53  & 2.30 & 2.60  \\
\br
\end{tabular}
\end{table}

\normalsize
The differences in the $U$ values associated with the underlying GGA or LDA functional are very small (0.2-0.3 eV) and can be neglected for all practical purposes. This also suggests that the $U$ is reasonably transferable across different functionals and pseudo-potentials. We have verified this conjecture by comparing ultrasoft pseudo-potential calculations with all-electrons LAPW methods \cite{blaha2001wien2k} with very satisfactory results.
Our ACBN0 results are validated against experiments and hybrid functionals, sX-LDA and  HSE, which incorporate some fraction of the non-local exchange-correlation effects and are, in spirit, similar to the ACBN0 functional.
The ACBN0 structural parameters are obtained by a full geometry optimization (cell and atomic degrees of freedom) and are tabulated in Table \ref{tab:tmo-lattice}. Overall, the results for the structural properties are in good agreement with experiments, and HSE and sX-LDA calculations. The ACBN0-LDA approach increases the lattice constant and brings the value closer to experimental data (see Table \ref{tab:tmo-lattice}). Our ACBN0-GGA calculations, on the other hand, lead to an expansion of the lattice. We remark that ACBN0-LDA describes the structural properties better than ACBN0-GGA as validated with respect to experimental results. 

\begin{table}[htb!]
\footnotesize
\centering
\captionsetup{width=\linewidth}
\caption{Equilibrium lattice constants $a_0$ in Angstrom calculated in this work and compared with various methods and experimental values. The stable AF2 magnetic ordering is considered}
\begin{tabular}{lcccc}
\mr
& MnO  &FeO & CoO & NiO \\
\hline
PBE& 4.52  & 4.39  & 4.30  & 4.31 \\
LDA & 4.42  & 4.26  & 4.17  & 4.12\\
PBE-ACBN0 & 4.58 & 4.42   & 4.33     & 4.19   \\
LDA-ACBN0& 4.49  & 4.35   & 4.28  & 4.15  \\
sX-LDA \cite{gillen2013accurate}& 4.33 & 4.27 & 4.32   & 4.23\\
HSE & 4.41\cite{archer2011exchange}   & 4.32\cite{meng2016density} & 4.20\cite{singh2016electronic}  & 4.18\cite{archer2011exchange} \\
Expt.\cite{archer2011exchange} & 4.44  & 4.33  & 4.26  & 4.17  \\
\br
\end{tabular}
\label{tab:tmo-lattice}
\end{table}
\normalsize

Fig. \ref{fig:TMO-band} shows the band structures of the four TM monoxides computed with 
ACBN0 (bottom panel in red) and simple PBE (top panel in black) functionals. All the band structures path shown here follow the AFLOW  standard \cite{curtarolo2012aflow}. Table \ref{tab:gaps} gives the values of the energy gap (E$_g$) as compared with previous theoretical calculations, experimental measurements, and all electron calculations with the mBJ exchange and correlation functional.  The results are summarized in Table \ref{tab:gaps}. The mBJ method \cite{tran2009accurate} is a semi-local exchange potential widely used to correct band-gaps in semiconductors. The mBJ band structures are plotted and discussed in the Supplementary Information (SI).
ACBN0 correctly predicts the four TM mono-oxides to be insulating with an indirect band-gap (see Fig. \ref{fig:TMO-band}). The valence band maximum (VBM) for MnO, CoO and NiO occurs at the Z point while for FeO, it is located between the Z and the L point. In all cases, the conduction band minimum (CBM) is  at the $\Gamma$ point. The CB lower manifold consists of the highly dispersive TM-\textit{4s} states and unoccupied TM-\textit{d} narrow bands.
The VB involves O-\textit{p} and occupied TM-\textit{d} states. 
Both MnO and NiO have a small gap within DFT (LDA/GGA) which is a consequence of the exchange and crystal-field splitting but is severely underestimated. 

In MnO, the ACBN0 correction pushes the \textit{3d} bands down in energy increasing their hybridization with the O-\textit{2p} states and opens the energy gap to 2.2 eV in good agreement with  hybrid functionals sX-LDA and HSE03. In NiO, PBE wrongly locates the \textit{4s} band above the unoccupied \textit{3d} bands, in disagreement with experiments\cite{li2005quasiparticle}.
The ACBN0 correction resolves this issue by correctly positing the CBM  at the $\Gamma$ point and leading to an indirect band gap of 3.8 eV.  The valence band width of 9 eV is  also in better agreement with experiments \cite{li2005quasiparticle} and advanced functionals. 
For both MnO and NiO, the mBJ calculations give a higher band-gap compared to the hybrid functionals. The mBJ electronic structure places the empty \textit{d} states of the Ni atom at the CBM which is at odds with experimental observations. 
The failure of PBE for both CoO and FeO is more dramatic since it predicts a metallic ground state. For these oxides, the energy gap should occur within the minority spin $t_{2g}$ levels. In CoO, with ACBN0 we calculate an energy gap of 3.25 eV in good  agreement with HSE calculations   with 25\% exchange. The valence band consists of  hybridized Co-\textit{3d}-O-\textit{2p} states, this is at odds to some XPS measurements \cite{qiao2013nature} which suggest that the Co \textit{3d} states are closer to the Fermi level and well separated from the dispersed O \textit{2p} states. The mBJ potential also yields a higher band-gap for CoO. 
FeO is less studied experimentally and few band gap measurements exist \cite{bowen1975electrical}. Theoretical studies with hybrid functionals  report a band-gap in the range of 2.5 to 2.8 eV \cite{heyd2003hybrid,aryasetiawan1998gw}.
 
\begin{figure*}[htb!]
\centering
\captionsetup{width=\linewidth}
\begin{tabular}{cccc}
\includegraphics[width=0.22\columnwidth]{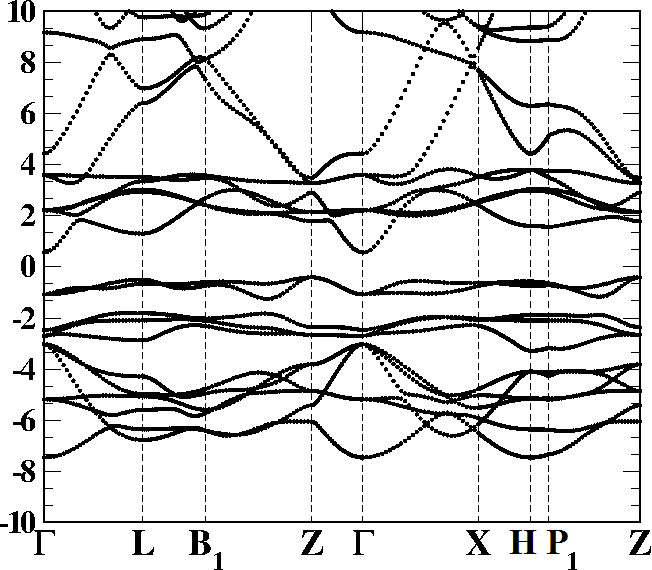} &  \includegraphics[width=0.22\columnwidth]{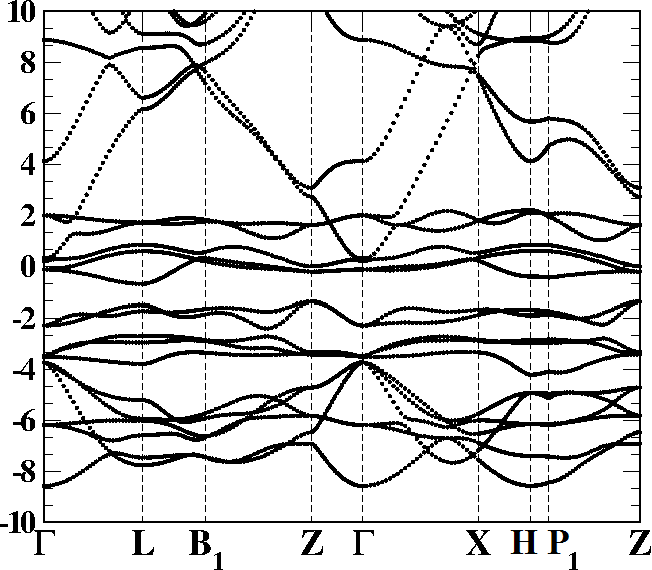} & \includegraphics[width=0.22\columnwidth]{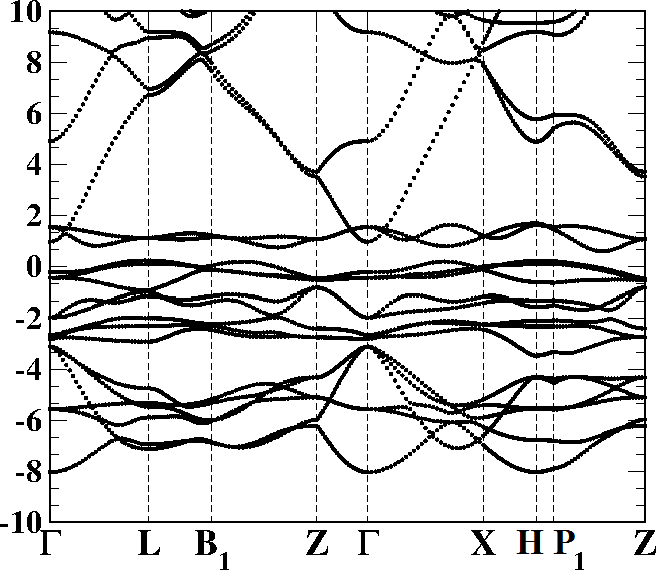} & \includegraphics[width=0.22\columnwidth]{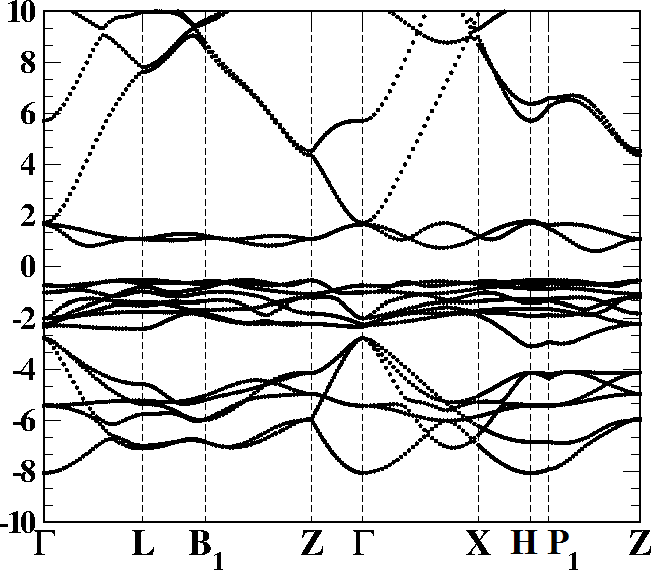} \\

\includegraphics[width=0.22\columnwidth]{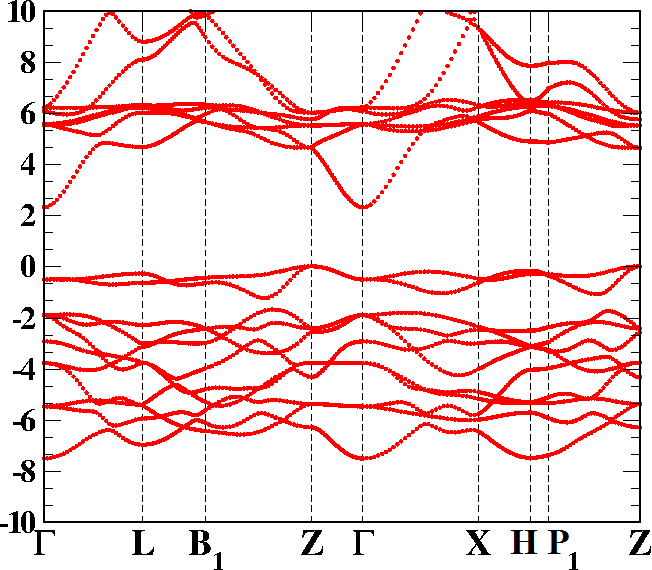} &  \includegraphics[width=0.22\columnwidth]{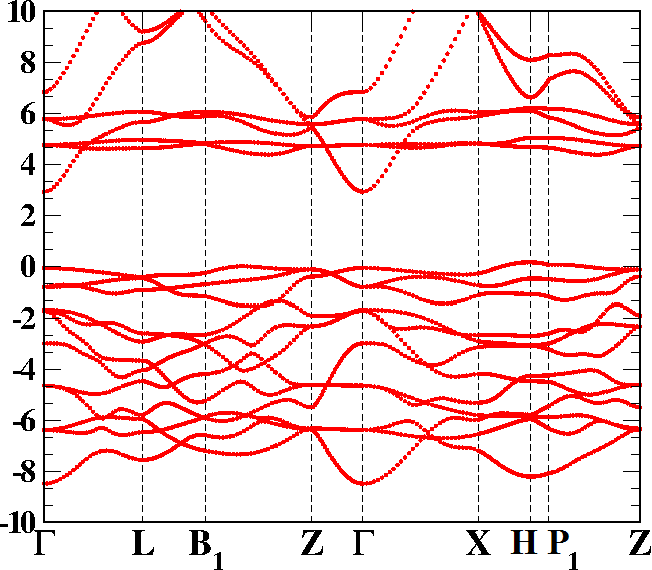} & \includegraphics[width=0.22\columnwidth]{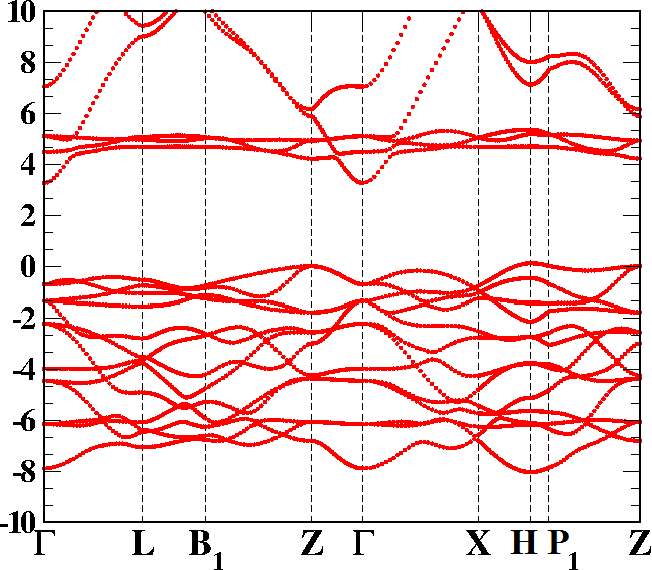} & \includegraphics[width=0.22\columnwidth]{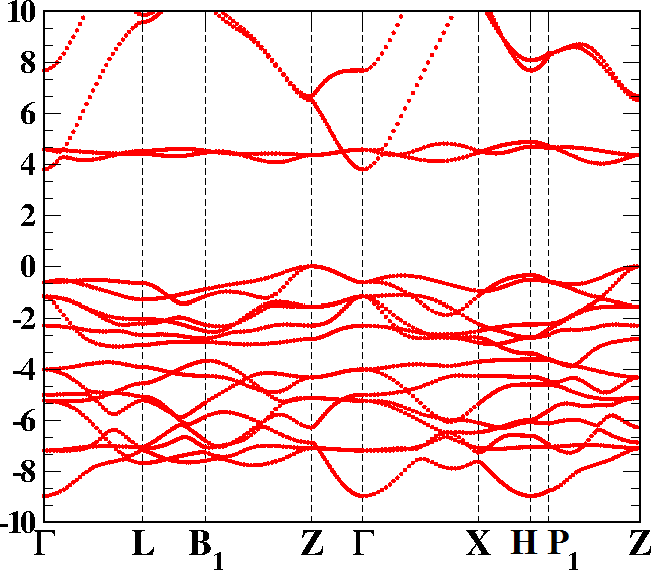} \\
a & b & c & d \\
\end{tabular}
\caption{\small{Band structure (spin up) of (a)MnO (b) FeO (c) CoO and (d) NiO. All energies are relative to the valence band maximum E$_\textrm{V}$. Effective values of $U$ for the TM \textit{d}-states and {O-$2p$} states as determined using ACBN0 in Table \ref{tab:Ueff-tmo} are used in the DFT+\textit{U} calculation. The top row are the PBE band structures.}}
\label{fig:TMO-band}
\end{figure*}

\begin{table*}
\centering
\captionsetup{width=\linewidth}
\footnotesize
\caption{\label{tab:gaps} Minimum direct and indirect energy bandgaps (in eV). The indirect band-gap is the Z-$\Gamma$ gap. In case of FeO, the VBM occurs at
a point between the B and Z symmetry point}
\begin{tabular}{lcccccccc}
\mr
& \multicolumn{2}{c}{MnO} & \multicolumn{2}{c}{FeO} & \multicolumn{2}{c}{CoO} & \multicolumn{2}{c}{NiO}  \\
& indir. & dir. & indir. & dir &indir.    &dir.&indir.              &dir. \\
\hline
PBE    &0.98 &1.64  &  \multicolumn{2}{c}{metallic} & \multicolumn{2}{c}{metallic}    & 1.13  & 1.26 \\
ACBN0  &2.31 & 2.83   & 2.70 & 2.86 & 3.25 & 3.95   &  3.80 & 4.30  \\
sX-LDA\cite{gillen2013accurate} &  2.5   & 3.0      & 2.3 & 2.4 & 2.7    & 3.7        & 4.04       & 4.3      \\
HSE03\cite{rodl2009quasiparticle}& 2.6\cite{schron2010energetic} & 3.2 & 2.1 & 2.2 & 3.2   & 4.0 & 4.1 & 4.5 \\ 
mBJ & 3.3    & 3       & \multicolumn{2}{c}{metallic} & 3.5    & 3.2        & 4.5        & 4.38     \\
Exp. (XAS-XES) &\multicolumn{2}{c}{4.1 \cite{kurmaev2008oxygen}} & & & \multicolumn{2}{c}{2.6 \cite{kurmaev2008oxygen}} & \multicolumn{2}{c}{ 4.0 \cite{kurmaev2008oxygen}} \\
Exp. (PES-BIS) & \multicolumn{2}{c}{3.9 $\pm$ 0.4 \cite{van1991electronic}}& & & \multicolumn{2}{c}{2.5\cite{van1991electronic}} & \multicolumn{2}{c}{4.3\cite{sawatzky1984magnitude}} \\      
Exp. (Conductance.& \multicolumn{2}{c}{3.8--4.2 \cite{drabkin1969photoconductivity}} &  &   & \multicolumn{2}{c}{2.5 \cite{van1991electronic2}} & \\
Exp. (absorption) & \multicolumn{2}{c}{3.6--3.8 \cite{iskenderov1969absorption}} & \multicolumn{2}{c}{2.4\cite{bagus1977width,balberg1978optical}} &\multicolumn{2}{c}{2.8\cite{gillen2013accurate}} &\multicolumn{2}{c}{4.0 \cite{gillen2013accurate}}\\
\br
\end{tabular}
\end{table*}
\normalsize

The magnetic moments of the TM atom are associated with the localized \textit{d} shells and are severely underestimated both in LDA and PBE (See Table.\ref{tab:magmom}). ACBN0 improves the  localization of the \textit{d} orbitals and brings the theoretical predictions in closer agreement with the experimental values for all the mono-oxides we studied with the exception of CoO. The measured moment in CoO includes both the spin and orbital contributions and hence is higher compared to our calculated ACBN0 values since we do not consider spin-orbit coupling. However, ACBN0 values are improved over the HSE values in both MnO and NiO and in good agreement for CoO and FeO.

\begin{table}[htb!]
\footnotesize
\centering
\captionsetup{width=\linewidth}
\caption{\label{tab:magmom} Local magnetic moments (in $\mu_B$) for the antiferromagnetic states of MnO, FeO, CoO and NiO}
\begin{tabular}{lllll}
\mr
   & MnO & FeO & CoO& NiO \\
\hline
LDA &   4.45   & 3.32   & 2.53 & 1.21   \\
PBE   & 4.58   & 3.44   & 2.60 & 1.49 \\
ACBN0-LDA  & 4.63 & 3.48 & 2.65 &1.72\\
ACBN0-PBE  & 4.79 & 3.59 & 2.8 & 1.83\\
HSE  \cite{rodl2009quasiparticle}    & 4.5  & 3.6 &2.7 &1.5\\
Expt. & 4.58 \cite{cheetham1983magnetic}, 4.79 \cite{fender1968covalency} &   3.32\cite{roth1958magnetic},4.2\cite{battle1979magnetic} & 3.35\cite{khan1970magnetic},3.98\cite{herrmann1978equivalent} &1.77 \cite{fender1968covalency}, 1.90 \cite{cheetham1983magnetic,roth1958magnetic} \\
\br
\end{tabular}

\end{table}
\normalsize

\bigskip
All the four TM mono-oxides are robust antiferromagnets with a N\'eel temperature ($T_N$) ranging from 100 K in MnO to 500 K in NiO. The phase transition between antiferromagnetic ordering and the corresponding paramagnetic phase can be described by estimating the exchange constants $J_{ij}$ in the Heisenberg Hamiltonian \cite{fischer2009exchange,gopal2004first}. Within DFT, the most common approach to evaluate $J_{ij}$ is to calculate the total energies of \textit{N} + 1 magnetic configurations where \textit{N} is the number of different exchange constants \cite{fischer2009exchange,archer2011exchange}.
We used this approach and calculate the total energies of three magnetic configurations (See Eqns. 6 and 7 of Ref. \cite{fischer2009exchange} for details on the formulas\ for $J_1$ and $J_2$). The ACBN0 values from Table \ref{tab:Ueff-tmo} are used for all the orderings. The magnitude of the exchange parameters are of the order of a few meV and the computations must be as free as possible from numerical errors. 

ACBN0 predicts the AF2 phase to be the most stable in all the four oxides in good agreement with experiments. The relative energies w.r.t the AF2 phases for the four cases are tabulated in (SI: Table S2). The values of the exchange couplings, $J_{ij}$, are in Table \ref{tab:exchange1}. In all the four TM mono-oxides, $J_2$  is negative with magnitude higher than the direct exchange coupling $J_1$ supporting a case for strong super-exchange. In FeO, both LDA and PBE wrongly describe the FM state to be most stable and hence we are unable to extract the $J_{ij}$ values. The ACBN0 correctly predicts the AFM stable state in FeO. Experimentally, the absolute value of $J_2$ increases in magnitude across the series from Mn to Ni. For NiO this value is the highest and is consistent with the higher N\'eel temperature observed experimentally \cite{potapkin2016magnetic}. 
The N\'eel temperature, $T_M$ is calculated by solving the Heisenberg Hamiltonian with the our computed \textit{ab initio} $J_1$ and $J_2$ values using a classical Monte Carlo (MC) simulation (Table \ref{tab:exchange1}). Additionally, we can also determine the Curie-Weiss temperature ($T_C$) which determines the susceptibility above the N\'eel temperature using  the expression $T_c = \frac{2}{3k_b}(6J_1+3J_2)S(S+1)$ \cite{harrison2007heisenberg}.

\begin{table*}[htb!]
\centering
\caption{\label{tab:exchange1}Nearest and next-nearest-neighbor exchange coupling constants $J_1$ and $J_2$ in meV. Comparison is made with different advanced functionals and experiments.}
\resizebox{\columnwidth}{!}{%
\begin{tabular}{lllll|llll|llll|llll}
\hline
\hline
    &\multicolumn{4}{c}{MnO}   & \multicolumn{4}{c}{NiO} &\multicolumn{4}{c}{CoO} & \multicolumn{4}{c}{FeO}\\
    &  $J_1$ & $J_2$ & Tn & Tc &  $J_1$ & $J_2$ & Tn & Tc & $J_1$ & $J_2$ & Tn & Tc & $J_1$ & $J_2$ & Tn & Tc \\
\hline
LDA &  -5.25      &  -13.18     & 246   &  769  &   2.96     & -40.07      & 900  & 1580 & -4.95   & -16.45   & 306  & 1807   &   &    &   &  \\ 
PBE &  -10.75          &  -14.50      & 277       & 1171      &   4.71     &-50.30 & 824 & 1897 &-3.57  & -13.62  & 254  & 2294  & &  &   & \\
ACBN0-LDA &  -4.81 & -5.82 & 98  &500  &  2.03 & -22.03 & 418 & 833 & -4.59 & -11.77   &  220 & 1825   & 1.48   & -8.97   & 220  & 900 \\
ACBN0-PBE &  -5.81      &   -3.88     & 38    & 507    &  0.80     &   -13.85 & 256 & 570 &  4.57   & -7.31  & 190 &1047   &1.68  &-12.41  & 270   & 1259\\
HSE \cite{archer2011exchange}&  -7.00      &  -7.8    & 125   & 783   & 2.3 &   -21.0    & 393     & 760 &  &   &   &   &   &    &   &      \\
SIC \cite{fischer2009exchange}& 1.36  &  -3.30     & 125   & 783   & 2.8&    -15.0   & 325     & 622 & 1.06 & -8.80 & 260 & 765 & 0.96 & -7.00 & 162 &     \\      
Expt.\cite{archer2011exchange}& -0.86      & -0.95      & 118   &  542  & 1.4   & -19.0      & 525   & 933 &0.70\cite{tomiyasu2006magnetic} &  -6.30\cite{tomiyasu2006magnetic} & 289\cite{fischer2009exchange} & 330 \cite{Kittel} & 1.84\cite{kugel1978low} &  -3.24\cite{kugel1978low} & 192\cite{fischer2009exchange} & 570\cite{Kittel} \\
\hline
\end{tabular}
}
\end{table*} 

\begin{figure}[htb!]
\footnotesize
\centering
\captionsetup{width=\linewidth}
\includegraphics[width=0.7\columnwidth]{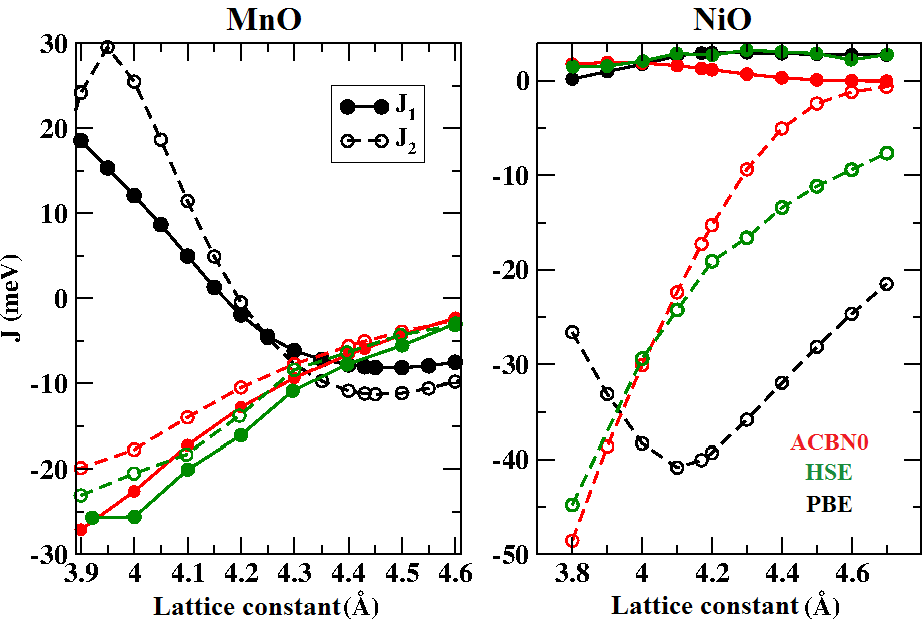}
\caption{\small{Magnetic exchange interaction for MnO and NiO at different lattice constants}}
\label{fig:mag-press}
\end{figure}

The variation of the magnetic properties with pressure is of critical importance and the ACBN0 approach has been used to investigate  the variation of exchange interactions as function of lattice constants. We restricted our investigation to MnO and NiO since they are prototypical systems and  can be directly compared  with hybrid functionals \cite{archer2011exchange}. Within ACBN0, \textit{U} depends on the geometry of the system\cite{gopal2015improved} and this is used to calculate the total energies and extract the exchange couplings, $J_{ij}$. Fig.\ref{fig:mag-press} shows the plot of the exchange interactions for a range of lattice constants for both MnO and NiO. The HSE data points are obtained from Ref.\cite{archer2011exchange} are in reasonable agreement with ACBN0 results.

Regular DFT (LDA/GGA) predicts positive  $J_1$ and $J_2$  in MnO under compressive strain, leading to a ferromagnetic metallic stable state. This incorrect description within DFT is corrected using ACBN0. 
In NiO, the relative ratio of $J_1$ and $J_2$ is predicted consistently by all functionals, however,  the magnitude of $J_2$, which is responsible for the magnetic ordering, is correctly reproduced only by HSE and ACBN0.
In both MnO and NiO, we see a remarkable agreement between ACBN0 and HSE values. 

\section{Results: mixed-valence Mn$_3$O$_4$ and Co$_3$O$_4$}
\label{sec:resultsMV}

Mn$_3$O$_4$ and Co$_3$O$_4$ are technological important materials in which the TM (TM=Mn,Co) atom has a mixed valence state due to the presence, in the spinel crystal structure, of both tetrahedrally (A) and octahedrally (B) coordinated sites.  The chemical formula can be written as A[B$_2$]O$_4$ with the A-site occupied by a divalent cation and B occupied by trivalent cations. According to \cite{robin1968mixed}, Mn$_3$O$_4$ and Co$_3$O$_4$ are  class II mixed valence compounds. The two TM atoms  have different charges and, hence, have different levels of correlation and exchange: they cannot be treated on an equal footing.  Unfortunately, there is not one single functional which can describe both the magnetic and the electronic properties accurately. For instance, recent studies by \cite{singh2014putting} showed that a single screening parameter ($\alpha=25\%$) in hybrid functionals could not reproduce the experimental results for both the electronic and magnetic properties. Within the ACBN0 approach the two TM sites can be treated independently leading to different values of the $U$ correction.

Co$_3$O$_4$ is a potential photovoltaic material with optical absorption in the visible range \cite{qiao2013nature,xie2009low}. It also exhibits interesting chemical and catalytic properties and therefore has great potential in novel renewable energy applications \cite{qiao2013nature}. 
Co$_3$O$_4$ crystallizes in the cubic normal spinel structure with space group \textit{Fd3m}, it is semiconducting with a band gap reported in the range 0.8-1.6 eV \cite{qiao2013nature,kim2003optical,waegele2014long}. Below $T_N \approx 40K $ the material is antiferromagnetic \cite{roth1964magnetic}. The magnetic structure of Co$_3$O$_4$ is relatively simple since the contribution to the magnetic moments derives completely from the A-site Co$^{2+}$ ions. The Co$^{3+}$ ions have no net magnetic moment (see Fig.1 of Ref. \cite{singh2014putting}).
Several groups have used a number of computational methods including PBE0, DFT+\textit{U}\cite{chen2011electronic}, HSE06 with varying values of $\alpha$ \cite{singh2014putting} to study the electronic and magnetic properties, yet the correct electronic structure and band gap is under debate. 
ACBN0 leads to a band gap of 1.2 eV and substantially rearrange the energy levels at the bottom of the CB (see Fig. \ref{fig:TMO-band}(a) and Table \ref{tab:div-exchange}). 
The improved accuracy is based on the ability of ACBN0 to treat differently the A and the B sites providing physically sound  $U$ values (Table \ref{tab:div-Uconv}).

As mentioned above, the magnetic exchange coupling for Co$_3$O$_4$ is dominated by the Co$^{2+}$ atoms on the A sublattice ($J_{AA}$). We determined the value for $J_{AA}$ by comparing the total energies of AFM and FM phases. Our results compare well with experimental reports as shown in Table \ref{tab:div-exchange}. 

\begin{figure}[htb!]
\centering
\footnotesize
\centering
\captionsetup{width=\linewidth}
\begin{tabular}{cc}
\includegraphics[width=0.3\columnwidth]{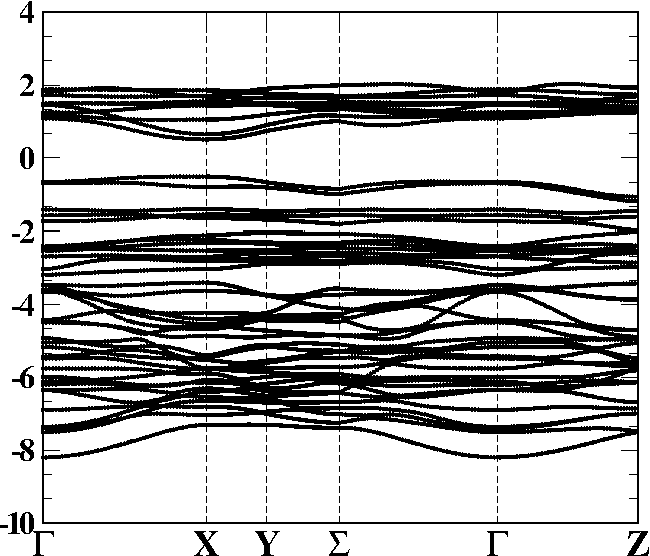} &  \includegraphics[width=0.3\columnwidth]{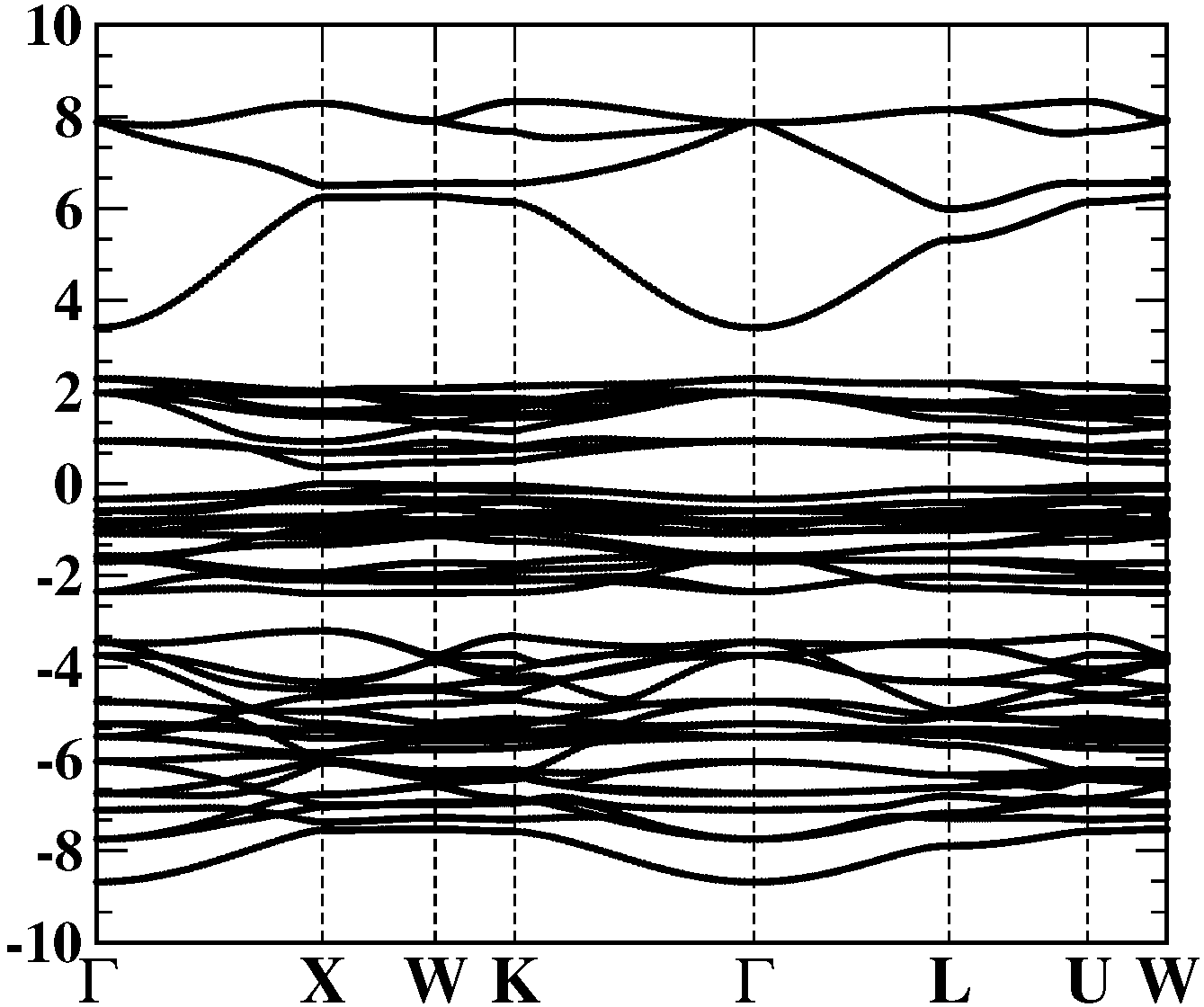} \\
\includegraphics[width=0.3\columnwidth]{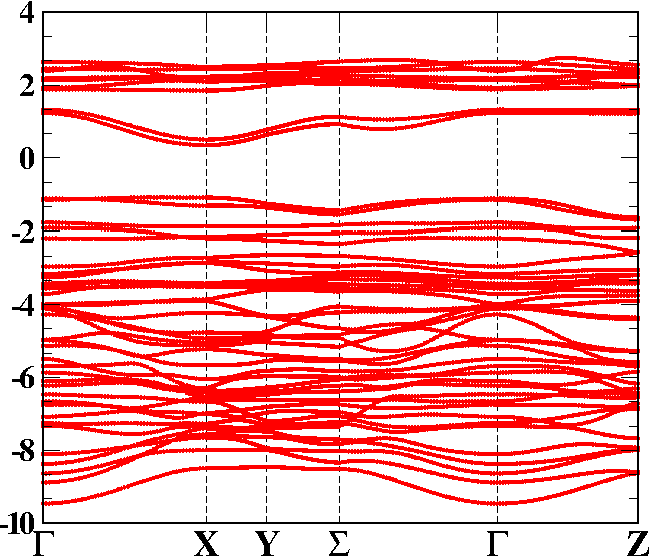} & \includegraphics[width=0.3\columnwidth]{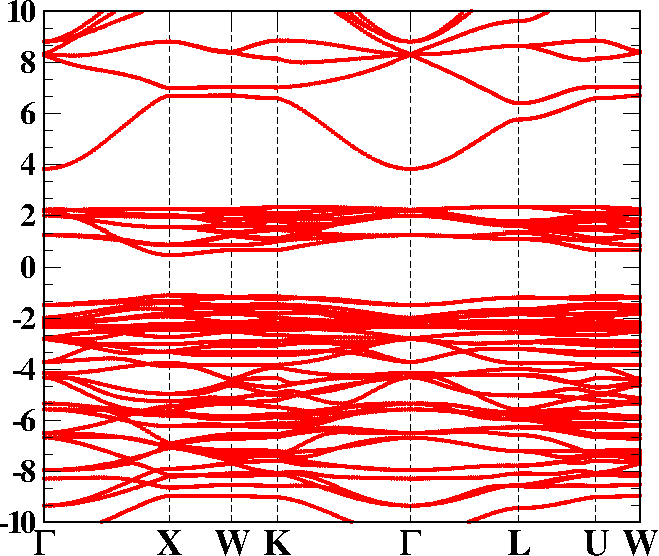} \\
a & b\\
\end{tabular}
\caption{Band structure of (a) Mn$_3$O$_4$ and (b) Co$_3$O$_4$ along the high symmetry points. Top panel shows the PBE band structures and the bottom panel shows the ACBN0 band structures.}
\label{fig:TMO-band}
\end{figure}
\normalsize

\begin{table}
\footnotesize
\centering
\captionsetup{width=\linewidth}
\caption{\label{tab:div-Uconv}Converged values of the effective on-site Coulomb parameter $U$ (in eV) for the transition metal (TM) $3d$ on the two sites and the oxygen $2p$ states.}
\begin{tabular}{lcc}
\mr
   & Mn$_3$O$_4$  &Co$_3$O$_4$ \\
\hline
TM-$3d$ (2+)      & 1.502 & 2.02 \\
TM-$3d$ (3+)      & 1.990 & 3.78  \\
Oxygen $2p$  & 3.735 & 3.49 \\
\br
\end{tabular}
\end{table}

\begin{table}[htbp]
\footnotesize
\centering
\captionsetup{width=\linewidth}
\caption{\label{tab:div-exchange} Atomic magnetic moment of Co(2+), electronic band-gap (in eV) and exchange coupling $J_{AA}$ between nearest neighbors in meV as calculated using PBE and ACBN0}
\begin{tabular}{cccc}
\mr
& ${\mu}_{Co(2+)}$  & E$_g$ (eV) &  $J_{AA}$ (meV)\\ 
\hline
\hline
PBE & 2.3  & 0.2 & -3.6  \\
PBE0\cite{chen2011electronic} & 2.90  & 1.96 &-5.0 \\
ACBN0 & 2.5 & 1.2 &-0.701\\
mbJ & 2.6 & 2.8 &  \\
HSE (25 pct.)\cite{singh2014putting} & 2.6 & 3.0 & -0.65  \\
HSE (5 pct.)\cite{singh2014putting} & 2.5 & 0.79 &-2.6  \\
Expt  & 3.25\cite{roth1964magnetic}   & 0.7\cite{qiao2013nature}, 0.9\cite{waegele2014long}, 1.6\cite{waegele2014long}, 1.65\cite{kim2003optical}  & -0.626\cite{scheerlinck1976magnetic} \\
\br
\end{tabular}
\end{table}

\normalsize

 Mn$_3$O$_4$ has a tetragonally distorted spinel structure (space group \textit{I4$_1$}, see Fig.1 of Ref. \cite{ribeiro2015density}). The experimental band gap in this oxide has been reported for thin films, E$_g$ = 2.51 eV  and  nanoparticles, E$_g$ = 2.07 eV \cite{hirai2015electronic}. To the best of our knowledge, no optical measurements for single crystal Mn$_3$O$_4$ were ever reported.
 For the lowest energy configuration (FiM6, see below), the band structure computed with ACBN0 slightly improves with respect to PBE calculations. The band gap increases from 0.3 eV (PBE) to  1.33 eV, closer to the reported experimental value of 0.9 eV \cite{hirai2015electronic}.
 
 In contrast to Co$_3$O$_4$, the magnetic structure of Mn$_3$O$_4$ is quite complex and undergoes three magnetic transitions and exhibits a non-collinear ferrimagnetic behavior \cite{hirai2015electronic,franchini2007ground}. In this work, we restrict ourselves to collinear magnetic couplings that capture the possible ferromagnetic/antiferromagnetic  interactions within the primitive spinel unit cell of 14 atoms. In the case of Mn$_3$O$_4$,  the two valence states (Mn$^{2+}$ and Mn$^{3+}$) contribute to the magnetic moment and leads to competing ground states. As a consequence, there may be different magnetic orderings  within a primitive unit cell of 14 atoms. The dominant exchange coupling constants are $J_{AA}$, $J_{BB}$ and $J_{AB}$ where A and B are the cations distributed in tetrahedral and octahedral sites respectively \cite{chartier1999ab,ribeiro2015density}.
We computed  the exchange interactions following Ref. \cite{chartier1999ab}  by approximating the experimental magnetic ordering of Mn$_3$O$_4$ using different collinear arrangements of the spin.
 Early theoretical work on this system includes a Hartree-Fock study of the exchange coupling constants whose results overestimate the experimental measurements \cite{chartier1999ab}. Ref. \cite{franchini2007ground} performed DFT calculations using various advanced functionals such as PBE+\textit{U}, HSE, and PBE0; they concluded that  hybrid functionals provide a better description compared to other methods. A more detailed comparison using the B3LYP, B3PW hybrid functionals were done in Ref. \cite{ribeiro2015density}. Within ACBN0, the \textit{U} values were calculated for the FiM6 configuration for the two different Mn atoms separately and these values were then used to calculate the total energies of all other configurations (Table \ref{tab:Mn3O4-energies}). 
ACBN0 results are improved with respect to PBE (Tables \ref{tab:Mn3O4-energies} and \ref{tab:Mn3O4-exchange}), however, some disagreement with experimental values remain. We speculate that there are many competing interactions and using a collinear model may not be appropriate for this system. 

\begin{table}[htb!]
\footnotesize
\centering
\captionsetup{width=\linewidth}
\caption{\label{tab:Mn3O4-energies} Total energy difference (\textit{$\triangle$E=E$_{FM}$-E$_{FiM}$}) in meV energies between different magnetic orderings in the primitive unit cell of Mn$_3$O$_4$ within PBE and ACBN0. For comparison, the values from two other functionals from Ref.\cite{chartier1999ab,ribeiro2015density} are also reported. The energy differences are given w.r.t to the FM configuration.}
\begin{tabular}{ccccc}
\mr
Configuration & PBE & ACBN0 &HF\cite{chartier1999ab} & B3LYP\cite{ribeiro2015density} \\
\hline 
FM &  0  & 0  & 0 & 0 \\
FiM1 & -688 &-254 & -66 & -108 \\
FiM2 &  -405 &-164 &-34 & -127\\
FiM3 &  -503 & -199 & -4& -217\\
FiM4 &   -786& -380 & -94&-356\\
FiM5 &   -440& -210 & -40&-110\\
FiM6 &   -838& -414 & -96&-402\\
\br
\end{tabular}
\end{table}
\normalsize

\begin{table}[htbp]
\footnotesize
\centering
\captionsetup{width=\linewidth}
\caption{\label{tab:Mn3O4-exchange} Different magnetic exchange couplings (in K) for Mn$_3$O$_4$ compared with other functionals and experimental values. }
\begin{tabular}{cccccc}
\mr
Coupling & PBE & ACBN0 &HF\cite{chartier1999ab} & B3LYP\cite{ribeiro2015density} & Expt\cite{srinivasan1983magnetic} \\
\hline
J$_{AA}$&  -65    & -7.77      & -4.13  & -5.73   & -4.9 \\
J$_{AB}$&  -33.2    & -0.902   & -3.11   & -20.01  & -6.8 \\
J$_{BB}$&  -35    & -15.5      & -6.57  & -20.52   & -19.9 \\
\br
\end{tabular}
\end{table}

\normalsize

\section{Summary and Conclusions}
\label{sec:Summary}

Results on the electronic structure and magnetic properties for several TM oxides were obtained using the new ACBN0 approach. We found great improvement with respect standard LDA/GGA DFT calculations. In particular ACBN0 capture the insulating character of FeO and CoO, adjusts the energy gap in MnO and NiO, and improve the overall band structure. The TM mono-oxides are correctly described as antiferromagnetic with local magnetic moments in agreement with experimental findings. The exchange couplings $J_1$ and $J_2$ are computed for MnO, NiO, CoO, and FeO are computed and systematically compared with LDA, PBE, HSE, SIC, and experimental data.

We also exploited the ability of the ACBN0 scheme to describe mixed-valence oxides, namely Mn$_3$O$_4$ and Co$_3$O$_4$. We demonstrated the dependence of \textit{U} from the oxidation state and improved the state of the art in term of band structure of mixed-valence oxides. Magnetic exchange couplings were also computed and analyzed.

\section*{Acknowledgements}

This work was supported by ONR-MURI under contract N00014-13-1-0635, DOD-ONR (N00014-14-1-0526) and the Duke University Center for Materials Genomics.
S.C. acknowledges partial support by DOE (DE-AC02-05CH11231, BES program under Grant \#EDCBEE).
We also acknowledge the Texas Advanced Computing Center (TACC) at the University of Texas Austin for providing HPC resources,
 and the CRAY corporation for computational assistance.
 
\newpage

\bibliographystyle{unsrtnat}
\newcommand{\Ozolins}{Ozoli\c{n}\v{s}}

\end{document}